# Design and Analysis of High Frequency InN Tunnel Transistors

**Krishnendu Ghosh and Uttam Singisetti**

*Department of Electrical Engineering, University at Buffalo, The State University of New York, Buffalo, NY 14260*
*Email: kghosh3@buffalo.edu , uttamsin@buffalo.edu*

**Abstract:**

This work reports the design and analysis of an n-type tunneling field effect transistor based on InN. The tunneling current is evaluated from the fundamental principles of quantum mechanical tunneling and semiclassical carrier transport. We investigate the RF performance of the device. High transconductance of 2 mS/µm and current gain cut-off frequency of around 460 GHz makes the device suitable for THz applications. A significant reduction in gate to drain capacitance is observed under relatively higher drain bias. In this regard, the avalanche breakdown phenomenon in highly doped InN junctions is analyzed quantitatively for the first time and is compared to that of Si and InAs.

# INTRODUCTION

Tunnel field effect transistors (TFETs) have achieved a lot of attention in recent years due to their superior subthreshold slope (SS) of below 60mV/decade thereby becoming a promising candidate for application in low power integrated circuits [1, 2]. III-V semiconductors due to their low effective tunneling mass and small band gap are efficient as a channel material for low power TFETs. However, their high speed radio frequency (RF) performance is not investigated much compared to the conventional MOSFETs. Low transconductance ($g_m$) and high gate to drain feedback capacitance ($C_{gd}$) are identified as the challenges to implement TFETs in high frequency applications [3-5]. Double gate (DG) TFETs with high-k dielectric [6] and gate all around (GAA) TFETs [7] have also been explored to improve the high frequency performance with limited success. Recently InAs vertical TFETs [8, 9] with an $n^+$ pocket in source have been demonstrated to boost up the on current ($I_{on}$) but high $C_{gd}$ is still an area of concern for TFETs as far as the RF performance is concerned.

InN is an attractive TFET channel material due to small electron effective mass ($0.04m_0$) and moderately high band gap (0.7 eV) which provides it potential for high off state breakdown voltage. This paper analyzes a single gate TFET based on InN which shows excellent RF performance with a current gain cut-off frequency ($f_t$) of around 0.5THz. The proposed device is simulated in SILVACO ATLAS [10] taking non local band to band tunneling model into account for better accuracy. The device energy band diagram obtained from simulation is used as a starting point for tunneling current calculation under different bias condition. The tunneling current calculation is done using WKB formalism taking into account conservation of transverse momentum. The DC simulation of the device is done and the simulated current is compared with the calculated current. The saturation of drain current and transconductance is discussed from carrier transport point of view

We also analyze the high frequency performance of the device. As reported in literature, high $C_{gd}$ in TFETs is a bottleneck to the RF operation of the devices. Usually TFETs are operated at a lower $V_{ds}$. This work shows that a higher $V_{ds}$ could be a solution to reduce $C_{gd}$ and hence to obtain an improved RF performance. But use of higher $V_{ds}$ makes it essential to explore the avalanche breakdown mechanism in the device. Avalanche breakdown in the proposed InN device is explored quantitatively. Finally we present the simulated small signal parameters under different bias and the trends of different parameters are explained from TFET electrostatics and carrier transport. The symbols used in this paper are tabulated in Table 1.

| Table 1: List of symbols used in this paper ||||
|---|---|---|---|
| Symbols used for tunneling action analysis | | Symbols used for avalanche action analysis | |
| Electric Field | $F$ | Electric Field in transport direction | $F_z$ |
| Wave vector (in different directions) | $k$ (with suitable subscripts) | Fermi wave vector | $k_F$ |
| Energy(in different directions) | $E$ (with suitable subscripts) | Electron Energy | $\epsilon$ |
| Oxide thickness | $t_{HfO_2}$ | Phonon energy | $E_R$ |
| Transistor body thickness | $t_{InN}$ | Bandgap | $E_g$ |
| Dielectric constant of InN | $\epsilon_{InN}$ | Ionization energy | $E_i$ |
| Dielectric constant of oxide | $\epsilon_{HfO_2}$ | Depletion width | $w$ |
| Screening length | $\Lambda$ | Electron mean free path | $\lambda$ |
| Current density | $J_{ds}$ | Ionization coefficient | $\alpha$ |

**ANALYSIS OF TUNNELING CURRENT**

*Device Structure:* The 50 nm gate length device structure used in the study is shown in Fig. 1(a). The channel doping of $10^{17}$ cm$^{-3}$ (n-type) is the lowest achievable in InN. A 3 nm thick HfO$_2$ is used for the gate dielectric. Other device parameters are listed in Table 2. It is noted that there are growth challenges for this structure particularly for the p-type layer. A polarization hole-doped structure [11] or a tunnel junction [12] could be used to overcome this. Here, we focus on

| Table 2: Design parameters considered for simulation and analysis | | | |
|---|---|---|---|
| *Oxide* | | *InN* | |
| Permittivity | 15.6 | Bandgap | 0.7 eV |
| Thickness | 3nm | Electron tunnel mass | $0.07m_0$ |
| Bandgap | 6 eV | Hole tunnel mass | $0.27m_0$ |
| *Gate Metal* | | *e* saturation velocity | $3\times10^7$ cm/s |
| Work function | 4.3 eV | Low field *e* mobility | 1500cm$^2$/V-s |
| *Source /Drain/Channel Doping* | | Body Thickness | 10nm |
| Source Doping | $10^{20}$ cm$^{-3}$ (p-type) | Gate length | 50nm |
| Drain Doping | $10^{20}$ cm$^{-3}$ (n-type) | Permittivity | 15.3 |
| Channel Doping | $10^{17}$ cm$^{-3}$ (n-type) [UID] | Electron Affinity | 5.34 eV |

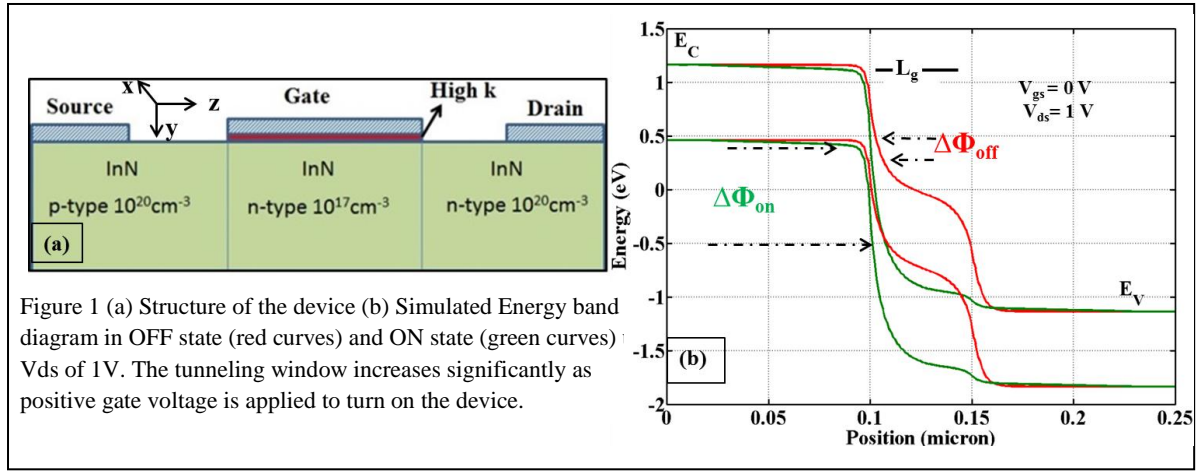

Figure 1 (a) Structure of the device (b) Simulated Energy band diagram in OFF state (red curves) and ON state (green curves) at Vds of 1V. The tunneling window increases significantly as positive gate voltage is applied to turn on the device.

fundamental electrical limitation of this device. Figure 1(b) shows the simulated energy band diagrams in ON and OFF states of the device.

*Tunneling Probability:* Using the simulated band diagram at a particular bias point, we calculate the tunneling probability under WKB formalism. We consider elastic tunneling; the total energy of the electron is same as before and after tunneling. We also take into account the conservation of transverse wave vectors since the potential barrier seen by the electrons is one dimensional only (in z direction) and the electron transport is also in that direction. The problem is simplified to a 1D tunneling problem incorporating the constancy of $k_x$ and $k_y$. Different values of $k_x$ and $k_y$ will give different longitudinal k-states ($k_z$) for a given total energy. Hence in our 1D model the transmission probability will be different for different values of transverse momentum $k_x$ and $k_y$. As depicted in Fig. 2, when an electron tunnels from the valence band (VB) to the conduction band (CB), due to the conservation of transverse wave vectors, the effective bandgap seen by a longitudinal k-state is given by [13]  $E_g^* = E_g + 2E_\perp$ where $E_\perp = \frac{\hbar^2(k_x^2+k_y^2)}{2m^*}$.

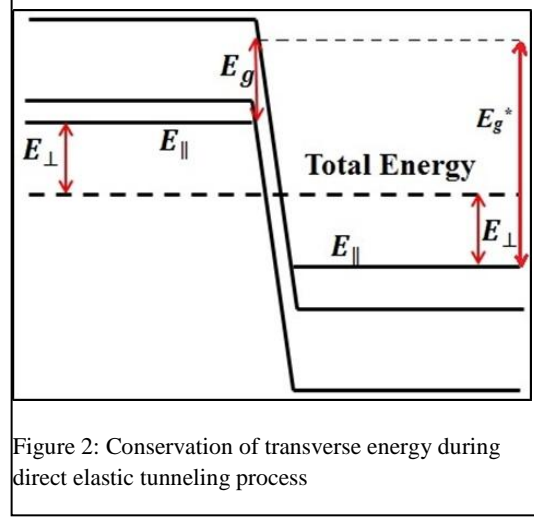

Figure 2: Conservation of transverse energy during direct elastic tunneling process

The transmission probability for a given longitudinal and transverse electron energy ($E_\parallel, E_\perp$) can be written as [14]

$$T(E_\perp) = \exp\left(-\frac{4\sqrt{2m^*}}{3q\hbar F}E_g^{*\frac{3}{2}}\right), \qquad (1)$$

where $F$ is the local electric field which is taken to be constant, $m^*$ is the reduced tunneling mass, $E_g^*$ is the effective band gap seen by the electron/hole while tunneling. The electric field is taken to be constant throughout the tunneling path. For better accuracy, F is taken to be 2/3 times (considering an exponentially decaying potential profile [15]) of the peak junction field, $F = \frac{2(E_g^* + \Delta\Phi)}{3\Lambda}$ where $\Delta\Phi$ is the tunneling window (Fig. 1(b)) which depends on the surface potential of the channel, $\Lambda$ is the screening length given by $\Lambda = \sqrt{\frac{\epsilon_{InN}}{\epsilon_{HfO_2}} t_{InN} t_{HfO_2}}$. Substituting the effective barrier and tunnel screening length in Eq. 1, expression for the transmission probability is

$$T(E_\perp) = \exp\left(-\frac{2\Lambda\sqrt{2m^*}}{q\hbar(E_g + 2E_\perp + \Delta\Phi)}(E_g + 2E_\perp)^{\frac{3}{2}}\right) \qquad (2)$$

The total tunneling probability is obtained by summing over the allowed transverse energy states. $T = \sum_{E_\perp} T(E_\perp)$. This summation can be converted to an integral [16] with the aid of two dimensional density of states in energy space. Next we calculate the allowed transverse states for a given longitudinal state $k_z$. The total energy should be within the tunneling window $\Delta\Phi$ on both sides of the junction; this introduces the constraint on allowed transverse states for given $k_z$. So $E_\perp$ can lie between 0 and $E_{max}$ where $E_{max} = \min(E_{vm} - E_\parallel, E_\parallel - E_{cm})$ $[E_\parallel = \frac{\hbar^2 k_z^2}{2m^*}]$. Here $E_{vm}$ and $E_{cm}$ are the minima of VB on the source side and maxima of CB on the channel side respectively. With this, the total tunneling probability for a given longitudinal is energy given by,

$$T(E_\parallel) = \int_0^{E_{max}} \rho(E_\perp) T(E_\perp) \, dE_\perp \qquad (3)$$

where $\rho(E_\perp)$ $(=\frac{\sqrt{m_e m_h}}{\pi\hbar^2})$ [16] is the two dimensional density of states in energy space. Figure 3(a) shows tunneling probability ($T(E_\parallel)$) normalized by $\int_0^{E_{max}} \rho(E_\perp) \, dE_\perp$) as a function of $E_\parallel$ which shows that as we move towards the band edges on the opposite sides of the source-channel junction ($E_c|_{channel}$ and $E_v|_{source}$) allowed values of $E_\perp$ decreases and hence $T(E_\perp)$ increases (from Eq. 2) which increases $T(E_\parallel)$. In the close vicinity of $E_c|_{channel}$ and $E_v|_{source}$ $T(E_\parallel)$ falls abruptly because $E_{max}$ becomes very small.

*Tunneling Current:* Now, with the effective bandgap to $E_g^*$ and the transverse energy states are taken care of within $T(E_\parallel)$, 1D tunneling current is given by,

$$J_{ds} = \frac{2q}{h}\int_0^{\Delta\Phi} T(E_\parallel)\left(f_L(E_\parallel) - f_R(E_\parallel)\right) dE_\parallel \tag{4}$$

where $f_L(E_\parallel)$ and $f_R(E_\parallel)$ are the Fermi distribution on two sides of the barrier. The factor $\frac{2q}{h}$ is the quantum conductance with opposite spins. The total current is obtained from the dimensions of the device as $I = \iint J_{ds} dA$, assuming the width of the device to be 1 micron.

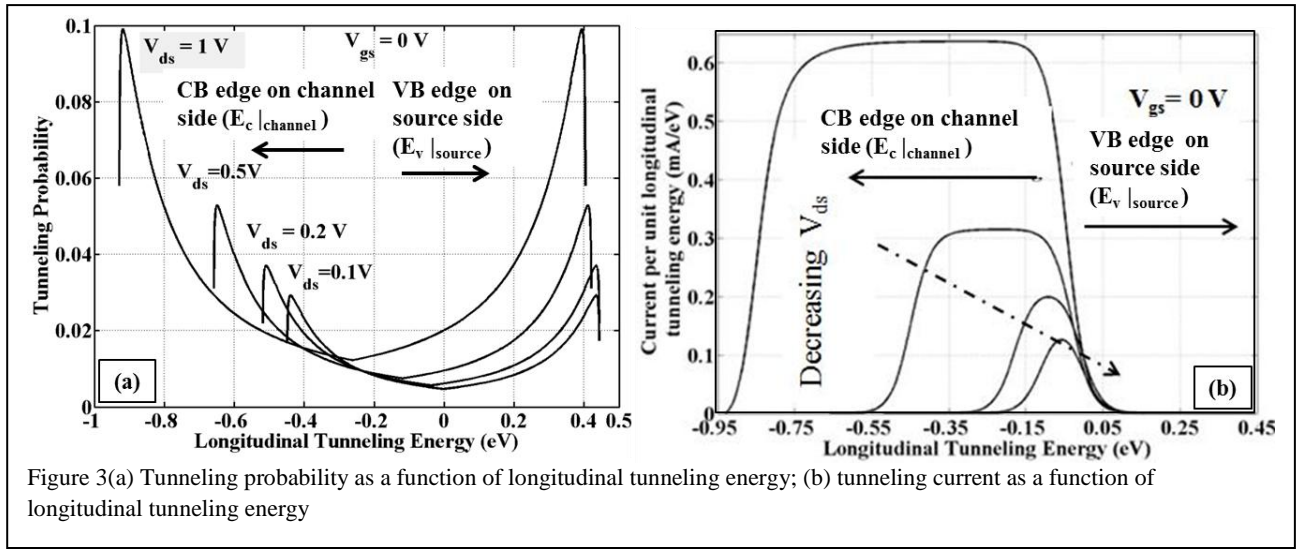

Figure 3(a) Tunneling probability as a function of longitudinal tunneling energy; (b) tunneling current as a function of longitudinal tunneling energy

The tunneling current contributed by a longitudinal state $E_\parallel$ as a function of $E_\parallel$ is shown in Fig. 3 (b). As we saw earlier $T(E_\parallel)$ falls abruptly in the close vicinity of $E_c|_{channel}$ and $E_v|_{source}$, the current also falls. Fig. 4 shows the calculated current matches the DC simulation results closely verifying the tunneling model in the simulator. The device can be well turned off at a gate voltage of -1V and gives an on/off current ratio of $4\times10^5$.

## DC SIMULATION

**Output Resistance:** The transfer characteristic is shown in Figure 4 while the output characteristic is shown in Figure 5(a) which looks quite similar to the conventional MOSFETs showing current saturation. The current saturation occurs due to the formation of a barrier from drain to channel at higher $V_{ds}$, this reduces the backward injection of electrons into the channel leading to saturation [15]. Unlike short channel MOSFETs the

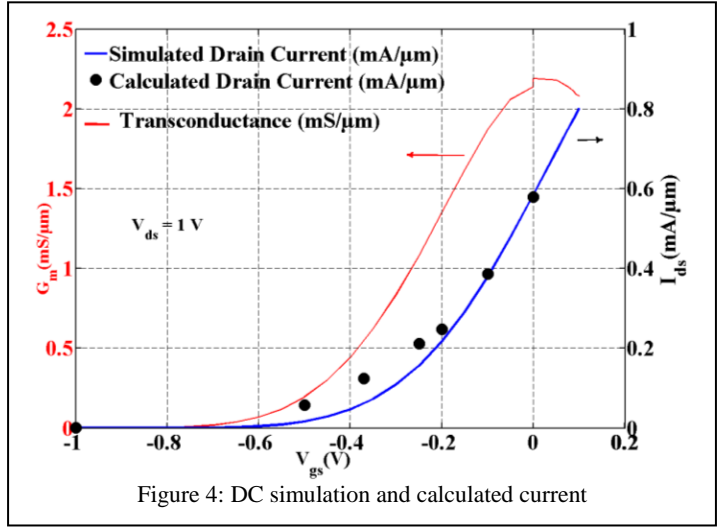

Figure 4: DC simulation and calculated current

dependence of $I_{ds}$ on $V_{ds}$ is much weaker for TFETs, the former is affected by the drain induced barrier lowering effect. For TFETs due to degenerate source doping the built-in potential at the source-channel junction ($p^+n^-$) at thermal equilibrium is so high that the VB edge on the source

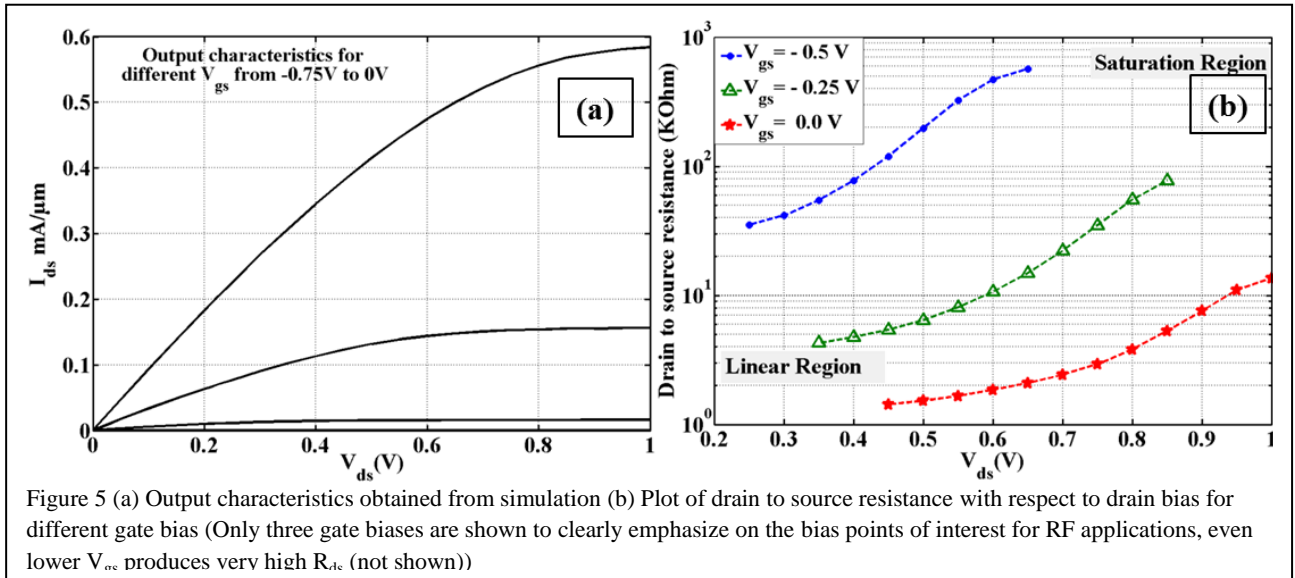

Figure 5 (a) Output characteristics obtained from simulation (b) Plot of drain to source resistance with respect to drain bias for different gate bias (Only three gate biases are shown to clearly emphasize on the bias points of interest for RF applications, even lower $V_{gs}$ produces very high $R_{ds}$ (not shown)) side is not effected after applying drain bias. The plot of drain to source resistance (Figure 5(b)) with respect to $V_{ds}$ shows the opposite trend to what is observed in short channel MOSFETs.

Here $R_{ds}$ increases with $V_{ds}$ because increasing drain bias increases the reverse bias of the channel-drain junction while in SC MOSFETs drain bias modulate the source-channel barrier, hence $R_{ds}$ drops with increasing drain bias. The saturating drain bias shifts to right as we choose a higher gate bias, because a more positive gate bias means a more positive drain bias is required to make the drain-channel barrier high enough to nullify the injection of electrons from the drain side.

## IMPACT IONIZATION ANALYSIS

**$C_{gd}$ and Avalanche breakdown:** As we discussed in the introduction the gate to drain capacitance ($C_{gd}$) is a bottleneck for RF operations of TFETs, next we discuss ways to reduce $C_{gd}$ of TFETs to get high RF performance. $C_{gd}$ originates from the injection of electrons from drain to gate which can be reduced if the barrier from drain to channel is high and the width of the depletion region is high. This can be achieved by high $V_{ds}$ but use of higher $V_{ds}$ poses a risk of avalanche breakdown in the drain-channel junction. In the following section, we show from gate electrostatics and impact ionization calculations that avalanche breakdown is less likely to happen under higher $V_{ds}$. Subsequently we do RF simulation of the proposed device under varying drain bias.

**Avalanche Mechanism:** The avalanche breakdown in a junction is accounted by the ionization coefficient ($\alpha$) and the junction breaks down if [13] $\int_0^w \alpha dx > 1$, where *w* is the width of the depletion region. Here we show that the InN TFET device can safely operate even at a $V_{ds}$ of 1V without breaking down leading to a drastic reduction in $C_{gd}$ and a significant increase in $g_m$ as required for high frequency performance.

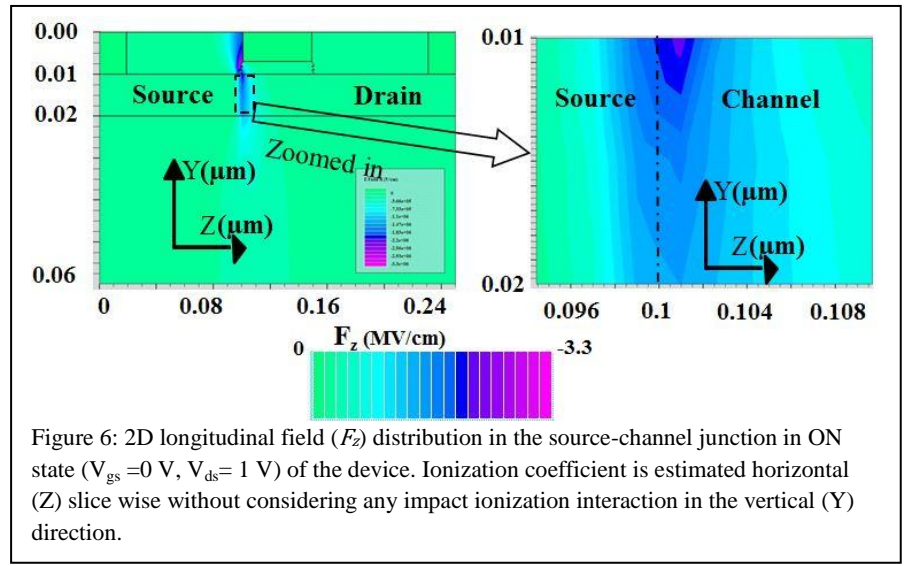

Figure 6: 2D longitudinal field ($F_z$) distribution in the source-channel junction in ON state ($V_{gs}$ =0 V, $V_{ds}$= 1 V) of the device. Ionization coefficient is estimated horizontal (Z) slice wise without considering any impact ionization interaction in the vertical (Y) direction.

When electrons traverse through the channel they can undergo collision with the lattice and can emit or absorb phonons. Due to the high longitudinal optical (LO) ($E_R$= 73 meV [17]) phonon energy in InN there are too few LO phonons present at room temperature to be absorbed by electrons. So electrons gain energy faster than energy loss by phonon emission. If the energy of an electron becomes equal to that of ionization energy ($E_i = 1.5E_g$ [14]) then the electron can cause impact ionization producing electron-hole pair. Thus the ionization coefficient depends on certain parameters like phonon energy ($E_R$), bandgap, mean free path $\lambda$ and the electric field in the junction. The electron mean free path for InN is calculated [18] to be $\lambda$ = 14.3 nm from $\lambda = \frac{3\pi^2 \hbar}{k_F^2 e^2 \rho_0}$; where $k_F$ is the Fermi wave vector, given by $k_F = (3\pi^2 n)^{\frac{1}{3}}$, $\rho_0$ is the resistivity and n is the electron concentration.

The two dimensional longitudinal electric field ($F_z$) distribution in the source-channel junction is shown Fig.6. There is a high field region near the junction, the electric field drops off moving away from the junction. The ionization co-efficient ($\alpha$) depends on the magnitude of electric field. Two different theories exist for two ranges of fields. We need to shift from one theory to the other when the field exceeds a threshold value given by [19] $F_{th} = \frac{E_R}{e\lambda}$. For InN, this is calculated to be 5×10$^4$ V/cm.

***High Field Regime:*** Under high fields beyond this threshold ($F_{th}$), the estimation of impact ionization coefficient can be done by Wolff's theory [20], $\alpha(F_z) \approx a \exp(-\frac{c}{F_z^2})$, where $F_z$ is the electric field. The constant c can be determined from the stationary distribution of electrons in energy space described by the distribution function [21], $f_0(\epsilon, F_z) \propto \exp(-\frac{\epsilon^2}{(eF_z\lambda)^2}\delta)$ where $\varepsilon$ is the electron energy, $\delta$ is the fraction of total energy lost due to phonon scattering. Now rate of impact ionisation is proportional to number of electrons with energy equal to ionization energy (i.e. $\alpha(F_z) \propto f_0(E_i, F_z)$). Hence, with $E_i = 1.5E_g$, we get, $\alpha(F_z) \propto \exp(-\frac{2.25E_g^2}{(eF_z\lambda)^2}\delta)$. Here $\delta(=\frac{E_R}{1.5E_g})$ is the fraction of total electron energy lost due to release of optical phonons. Hence the constant c for InN is calculated as $c = \frac{1.5E_RE_g}{(e\lambda)^2} = 0.037$ MeV²/cm² . The constant $a$ in the expression for α can be determined from experimental data. Baraff [19] has performed numerical calculations to plot α for different values of $\frac{E_R}{E_i}$. For InN, we have $\frac{E_R}{E_i} = 0.0695$, using this in the Baraff's plot the value of $a$ is evaluated as, $a = 0.01$ nm⁻¹. Under high field approximation, which is the dominant case for TFET operation, the expression for ionization coefficient for an InN is given by

$$\alpha(E) = 0.01\exp\left(-\frac{0.037}{F_z^2}\right) \qquad (5)$$

where $F_z$ and α are expressed in MV/cm and nm⁻¹ respectively. Similar calculation is done for InAs and Si and all the three curves are plotted in Fig. 7. The curve for Si matches closely with the measured one given in [14]. Thus the validity of the approximation is confirmed.

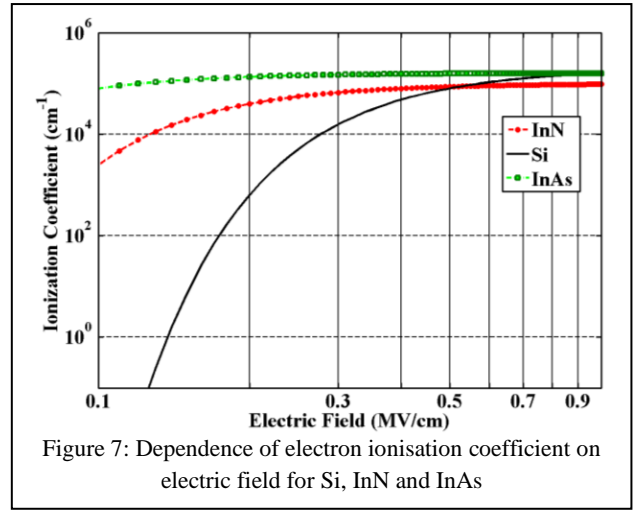

Figure 7: Dependence of electron ionisation coefficient on electric field for Si, InN and InAs

*Low Field Regime:* Under low fields below the threshold value, according to Shockley's theory [22], $\alpha(F_z) \approx a \exp(-\frac{E_i}{eF_z \lambda})$. Here the constant *a* can be determined experimentally but is not known for InN and cannot be determined from Baraff's plot either as the curve for InN ($\frac{E_R}{E_i} = 0.07$) doesn't extend till the low field regime ($F_z < 5 \times 10^4$ V/cm). So *a* is taken to be 0.07 nm$^{-1}$ (value for silicon [23]) which could overestimate α (since $\frac{E_R}{E_i}|_{Si} < \frac{E_R}{E_i}|_{InN}$ [14]).

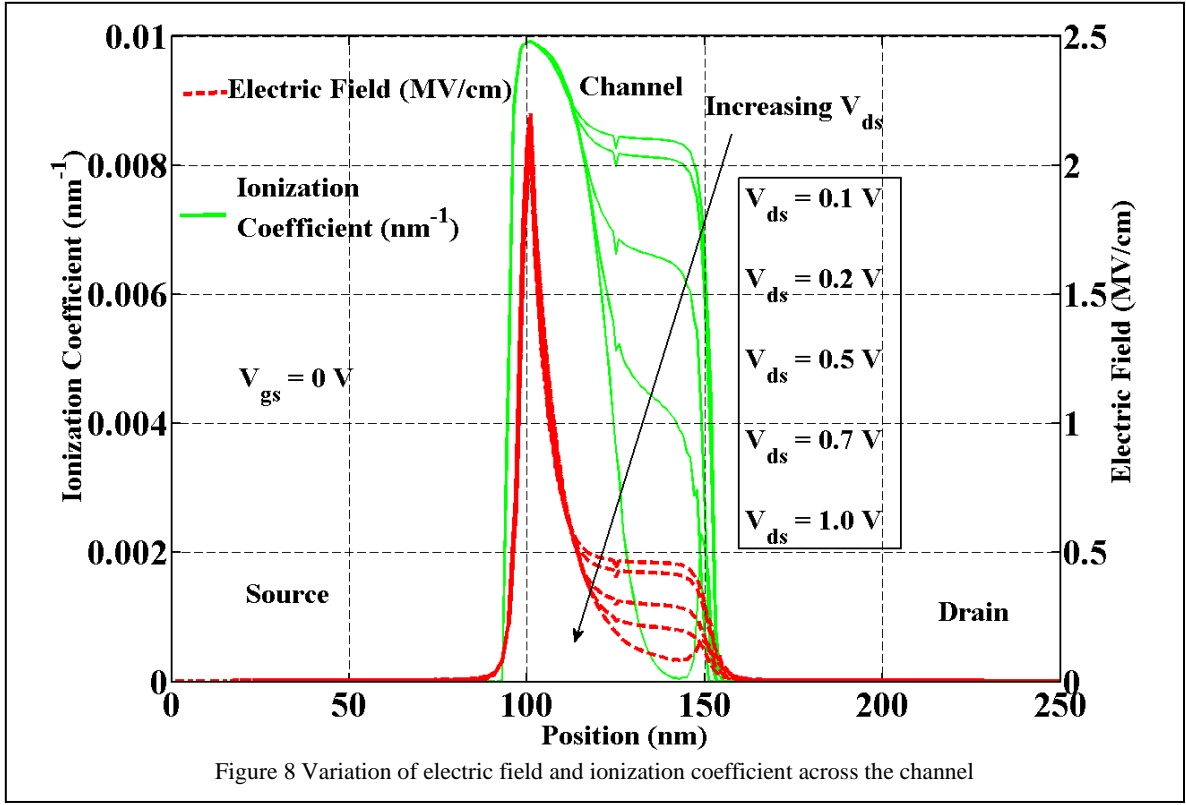

Figure 8 Variation of electric field and ionization coefficient across the channel

| Table 3: $\int_0^w \alpha dx$ for different drain bias ||
|---|---|
| $V_{ds}$ (V) | $\int_0^w \alpha dx\|_{InN}$ |
| 0.1 | 0.4967 |
| 0.2 | 0.4842 |
| 0.5 | 0.4265 |
| 0.7 | 0.3625 |
| 1.0 | 0.2804 |

The ionization coefficient is calculated for InN TFET under different drain bias conditions. The electric field and calculated ionization coefficients are shown in Fig. 8, and the calculated ionization integral $\int_0^w \alpha dx$ is given in Table 3. From Table 3 it can be seen that the value of the integral is counter intuitive as it seems like under high drain bias the device is less likely to undergo avalanche breakdown. But actually under low drain bias the field across the channel in ON state is much higher compared to that under high drain bias because under high drain bias there is less number of charges injected from the drain side near the channel/drain interface. This helps increase the surface potential [15] of the channel near the drain interface; which means the electric field will be less in the channel under high drain bias. Similar analysis for an InAs TFET gave $\int_0^w \alpha dx \sim 1$ at 1V $V_{ds}$. The band gap and optical phonon energy both are higher for InN, which makes InN TFETs more robust against such on state breakdown phenomenon.

## SMALL SIGNAL SIMULATION

Having established the safe operation at higher $V_{ds}$, we evaluate the small signal performance under different $V_{ds}$. The simulator applies a sinusoidal perturbation over the given

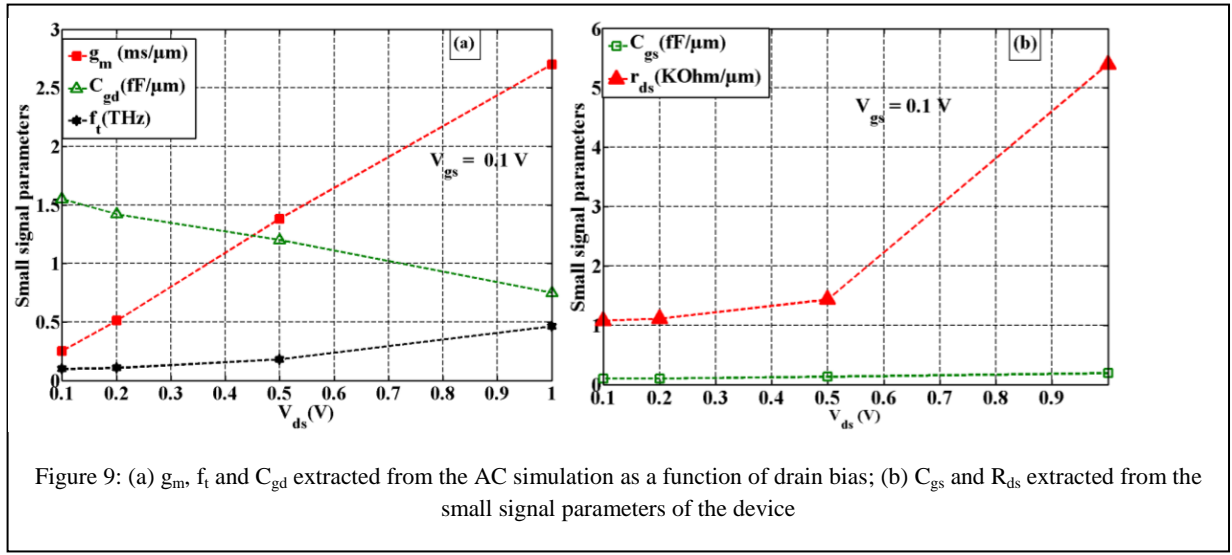

Figure 9: (a) $g_m$, $f_t$ and $C_{gd}$ extracted from the AC simulation as a function of drain bias; (b) $C_{gs}$ and $R_{ds}$ extracted from the small signal parameters of the device

DC bias point. Then it estimates the transient response with an infinite time boundary. The admittance matrix (Y parameters) is determined from the transient current and voltage outputs [24]. The real and imaginary components of the Y parameters yield the conductance and capacitance respectively between corresponding ports. The extracted small-signal parameters from the simulated y-parameters are plotted in Figs 9(a, b). As discussed before $C_{gd}$ is found to decrease (Fig. 9(a)) with increasing drain bias due to higher drain to channel barrier. In addition the transconductance of the device increases with drain bias. The decreasing $C_{gd}$ and increasing $g_m$ together boost up $f_t$. The device gives a peak current gain cut-off frequency of 460 GHz (Fig. 10) at 1V drain bias. The $C_{gs}$ (Fig. 9(b)) remains almost unchanged with drain bias, which can be explained as follows. The slight increase in $C_{gs}$ occurs because of the increasing tunneling probability with respect to drain bias. On the other hand, the significant increase in current (and hence in transconductance) occurs because of the simultaneous effect of tunneling probability increase and a reduction of electrons with negative $k_z$ states (due to higher drain to channel barrier) in the channel due to increased $V_{ds}$. The parasitic source resistance in a FET will degrade the high frequency performance. However, it is easier to form low resistance contacts to InN [25], which will not degrade the simulated high frequency performance.

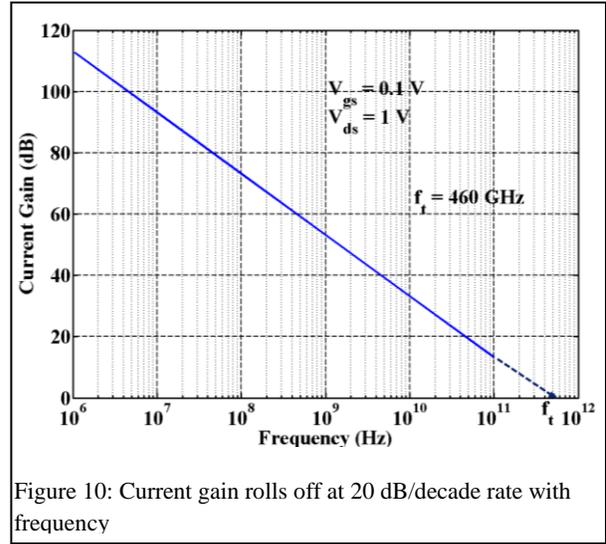

Figure 10: Current gain rolls off at 20 dB/decade rate with frequency

## CONCLUSION

In summary, the operation of an InN TFET is discussed with suitable simulation results for the first time. It has been shown that a relatively less gate to drain capacitance and hence a higher $f_t$ can be achieved at higher drain bias ($V_{ds}$ =1 V). It has also been quantitatively proved that that the device does not undergo avalanche breakdown at high drain bias. The relation between ionization coefficient and electric field for InN devices is also explored which can be useful in future for development of other InN devices as well. Finally a TFET design is proposed which has the potential of high power THz applications and at the same time gives good on/off current ratio for applications in CMOS logic circuits.